\documentclass[10pt,letterpaper]{article}
\usepackage{opex3}
\usepackage{graphicx}

\begin{document}

\title{Planar wallpaper group metamaterial for novel terahertz applications}

\author{Christopher M. Bingham$^1$, Hu Tao$^2$, Xianliang Liu$^1$, Richard D. Averitt$^3$, Xin Zhang$^2$, and Willie J. Padilla$^1$}

\address{
$^1$Department of Physics, Boston College, 140 Commonwealth Ave.,
Chestnut Hill, Massachusetts 02467, USA.\\
$^2$Boston University, Department of Manufacturing Engineering, 15
Saint Mary's Street, Brookline, Massachusetts 02446,
USA.\\
$^3$Department of Physics, Boston University, 590 Commonwealth
Avenue, Boston, Massachusetts 02215, USA. }

\email{binghach@bc.edu}

\begin{abstract}
We present novel metamaterial structures based upon various planar
wallpaper groups, in both hexagonal and square unit cells.
An investigation of metamaterials consisting of one, two, and three
unique sub-lattices with resonant frequencies in the terahertz (THz)
was performed. We describe the theory, and perform simulations and
experiments to characterize these multiple element metamaterials. A
method for using these new structures as a means for bio / chemical
hazard detection, as well as electromagnetic signature control is
proposed.
\end{abstract}



\section{Introduction}

Over the last several years the field of metamaterials has seen
enormous and consistent growth. Initially this was due to the demonstration of
negative refractive index (NRI), or ``left-handed materials" as it
was termed \cite{veselago68,smithPRL00,shelbySci01}, as well as the
prediction and verification of the ``perfect lens" \cite{pendry00,grbicPRL04}. This lead to strong interest from
scientists in many varied areas and spawned numerous major research
thrusts thus engaging a worldwide community. Metamaterials received
another enormous boost recently when it was discovered that it is
possible to construct a cloak - an advanced structure that makes an
object ``invisible" to interrogating light of a particular
wavelength \cite{pendrySci06,schurigSci06}. Present excitement in
metamaterials stems from their ability to access regimes of material
response not possible with naturally occurring materials. They also serve as a
platform to explore new physics in an otherwise mature science field
- classical electricity and magnetism.

NRI and cloaking are certainly booming areas of scientific
exploration and largely responsible for the immense growth of
metamaterials. However, artificial materials represent a new
paradigm for the construction of novel devices, and metamaterials
are much broader than just the examples listed above. Arguably, the
real power in metamaterials lies in the fact that it is possible to
construct materials with an exact magnetic and electric response.
For example, the so-called ``split ring resonator"~\cite{pendry99}
has been the standard metamaterial utilized for magnetic response
across much of the electromagnetic (EM) spectrum. With SRR type
structures demonstrated at radio~\cite{Wiltshire2001},
microwave~\cite{smith00,schurigAPL06}, mm-Wave~\cite{ekmel},
THz~\cite{science04}, MIR~\cite{soukoulis}, NIR~\cite{brueck,shalaevNIR} to
the optical~\cite{optical,shalaevopt} the design has proven to be very
versatile. The majority of prior metamaterials work has used simple
variations of the split ring design and/or straight wire media.

Although the above listed successes of metamaterials evince their
ability to access exotic EM responses and highlight their great
technical potential, they still face several major obstacles which
must be overcome before widespread usage in devices is possible. For
example, researchers strive to construct metamaterials such that
they may be described by bulk material parameters characterizing
their frequency dependent effective electric $\epsilon(\omega)$ and
effective magnetic $\mu(\omega)$ EM responses. As such, most designs
are based on resonances in order that the physical dimensions of the
unit cell ($a$) and the metamaterial itself ($L$) are smaller than
the wavelength at resonance $\lambda_0$, i.e. $L < a << \lambda_0$.
This resonant behavior is, of course, what permits metamaterials to
attain electromagnetic responses which deviates from unity - the
free space values - thereby achieving the exotic behaviors noted
above. However, a consequence of this strong frequency dispersion is
that the novel responses that metamaterials achieve are often narrow
band, typically $<$5\%. The combination of wire media and SRRs(along with
other similar variants) has also yielded great triumphs,
however these designs will still be limited by narrow band behavior and the metamaterial responses
achievable through combination. An important step in the
creation of artificial materials is to create designs that can mimic
the many various types of symmetries that are observed in nature.

We present new metamaterial geometries which may be oriented in
various different planar wallpaper groups that significantly extend
the pallet of available structures beyond simple square
lattices of wire and SRR arrays. Further, these new metamaterials
may help to alleviate the narrow-band response of metamaterials by
permitting multi-frequency response without significantly
compromising ``electromagnetic strength". We detail the theory,
simulation, fabrication, and measurement of several of these
metamaterials at THz frequencies. Several potential applications of
these new MMs are discussed including electromagnetic signature
control and biodetection.

\section{Theory}

The analogy between metamaterials and real materials can be
extended, beyond what has been shown before, by appealing to designs
that nature has provided. For example, condensed matter describes
materials consisting of different elements as a Bravais lattice with
a basis. In a similar manner, we can form metamaterials that have
more than one primitive cell, where each has a distinct metamaterial
element, and distinct electromagnetic properties. These individual
cells may then be added together to form a lattice which preserves
the electromagnetic properties of each sub-lattice. In our
particular study we investigate planar geometries and thus consider
various wallpaper groups \cite{wallpaper}. For example, if we are interested in
constructing a material with two distinct elements (bipartite), then
a planar MM array can be generated in a straightforward manner, e.g.
a checkerboard pattern. This has already been achieved
in efforts to eliminate bianisotropy in planar SRR
arrays~\cite{srr3} and to create dual resonance behavior~\cite {dual}.
However this arrangement may be less
desirable when moving to three elements (tripartite) or more. This
is due to the fact that the most effective geometry is one that will
allow for maximum filling fraction in order to maximize the EM
response, i.e. oscillator strength~\cite{pendry99}. Thus we should
appeal to unit cell symmetries different than square if we are
interested in achieving metamaterials with three or more distinct
elements. Considering only two-dimensional designs there are limited
primitive cell choices, and only parallelogram, triangular, and
hexagonal shapes can tile the whole surface - utilizing translation,
reflection, and rotational symmetries - without leaving gaps
in-between. Most metamaterial designs to-date use square lattices -
a special case of parallelogram\cite{shalaevnote}.

\begin{figure}
\begin{center}
\includegraphics[width=5in,keepaspectratio=true]{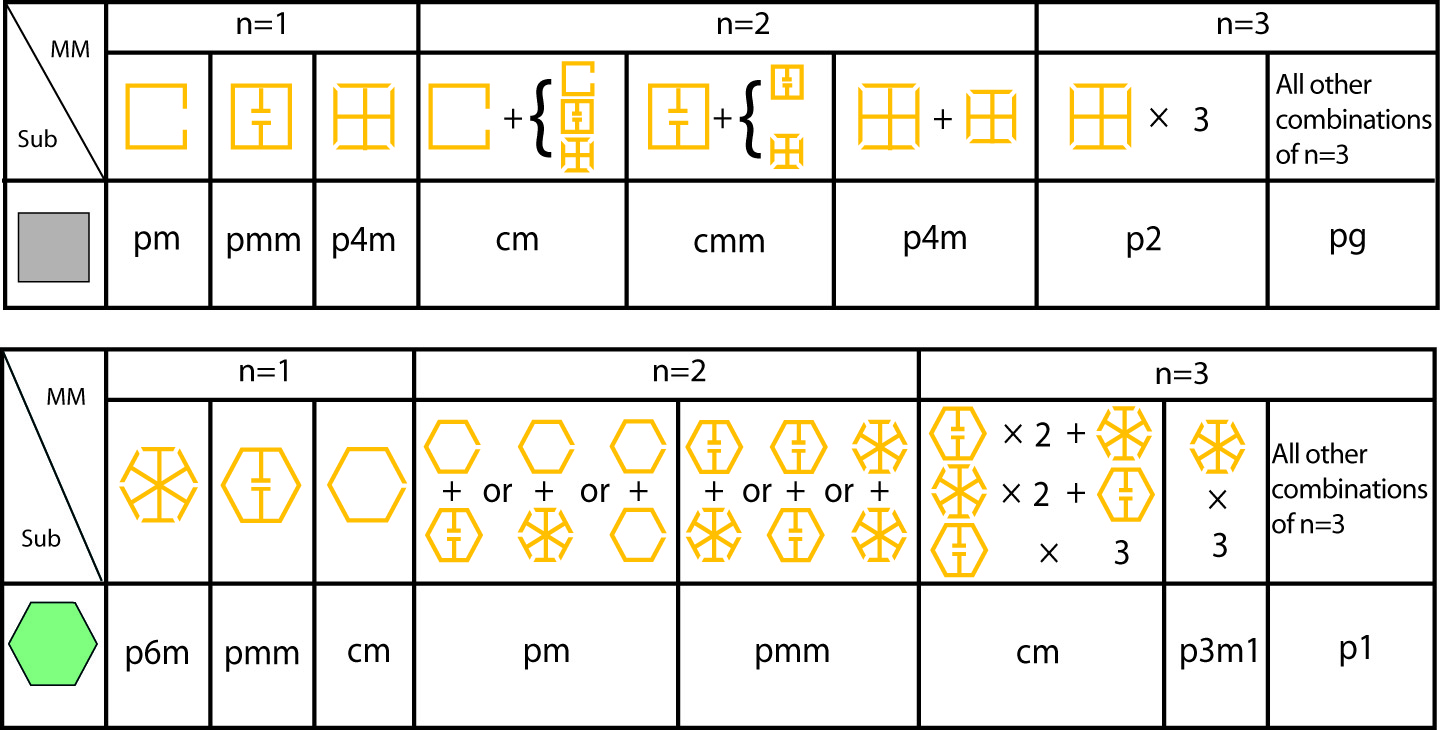}%
  \caption{Group classification for two dimensional metamaterial designs. }\label{table1}
  \label{fig1}%
  \end{center}
\end{figure}

Fig. \ref{fig1} shows the group classification for two dimensional
metamaterial designs using either square or hexagonal unit cells.
Appealing to the formalism of two dimensional wallpaper groups, the
top portion of Fig. \ref{fig1} gives all the possible groups
resulting from the combination of square primitive cells up to n=3
with different structures. For example, by placing one of the three
square MMs together (under the n=1 heading) and tiling the entire
surface, the pattern belongs to the group $pm$, $pmm$ or $p4m$.
Columns under the n=2 heading show groups with two different MMs in
one unit cell. In this case the MMs may tile the surface as a
checkerboard pattern. The last n=3 columns give groups that are
combinations of three primitive cells spanned by two tiling vectors
are along -30 and 30 degrees. We consider the case that if there is
more than one primitive cell per structure, that MMs with the same
symmetry are distinct, i.e. they are resonant at different
frequencies.

The bottom part of Fig. \ref{fig1} shows groups that correspond to
hexagonal primitive cells and MMs. A single hexagon shape and its
tiling to the entire two dimensional surface belongs to the group
$p6m$. This group is of highest symmetry among all the 17 wallpaper
groups, and has rotations of order $s$ = 2, 3 and 6, (where we
define a rotation as $2\pi/s$ radians), as well as various
reflection symmetries. Adding any lower symmetry object to the
hexagon will reduce the symmetry, as detailed under the n=1 heading.
If the number of hexagons in one primitive cell continues to
increase, the symmetry of the two dimensional pattern will keep
decreasing. For unit cells composed of n=2 MMs, the tiling vectors
are along 0 and 60 degrees. Unit cells consisting of n=3 MMs, (with
the three primitive cells being nearest neighbors), the tiling
vectors are along 0 and 60 degrees. Here, as with square unit cells,
we consider hexagonal MMs with the same symmetry to consist of
unique designs. When the number $n\geq$5 there will be no rotational
or reflection symmetry at all, belonging to group $p1$, which is the
group with the lowest symmetry in two dimensional wallpaper groups.

For a given design consisting of one primitive cell, the filling
fraction depends on the ratio of the area enclosed by metamaterial
to the area of the unit cell \cite{pendry99}. The frequency
dependence of the metamaterial response functions is given as:
\begin{equation}
\widetilde{\epsilon}(\omega),\widetilde{\mu}(\omega)=\epsilon_\infty,\mu_\infty+\sum_{m}\frac{F_{m}{\omega}^{2}}{{\omega}_{0m}^{2}-{\omega}^{2}-i\Gamma_{m}\omega}
\end{equation}

\noindent where $\omega_0$ is the center frequency of the m$^{th}$
oscillator, $\mu_\infty$, $\epsilon_\infty$ are the real parts of
$\mu$,$\epsilon$ at $\omega=\infty$, and $\Gamma$ is the loss. The
term $F_m$ is the filling fraction of the m$^{th}$ oscillator and is
given as,

\begin{equation}
{F_m}=\frac{a_m}{nA}
\label{fill}
\end{equation}

\noindent where $a_m$ is the area of the $m^{th}$ metamaterial
element, $n$ is the total number of distinct primitive cells, and
$A$ is the area of the primitive cell. The exotic behavior of
metamaterials, e.g. negative refraction, is obtained by maximizing
the element response such that negative parameters are obtained. As
such it is important to make the unit cell and the metamaterial
element the same shape, i.e. both square, or both hexagonal, in
order to maximize the filling fraction, and thus the electromagnetic
response. Increasing the number of primitive cells with different
elements will sharply decrease $F_m$, roughly dividing by the number
of primitive cells in a given design, as shown in Eq. \ref{fill}.

 \begin{figure}[ptb]
\begin{center}
\includegraphics[ width=5in,keepaspectratio=true]{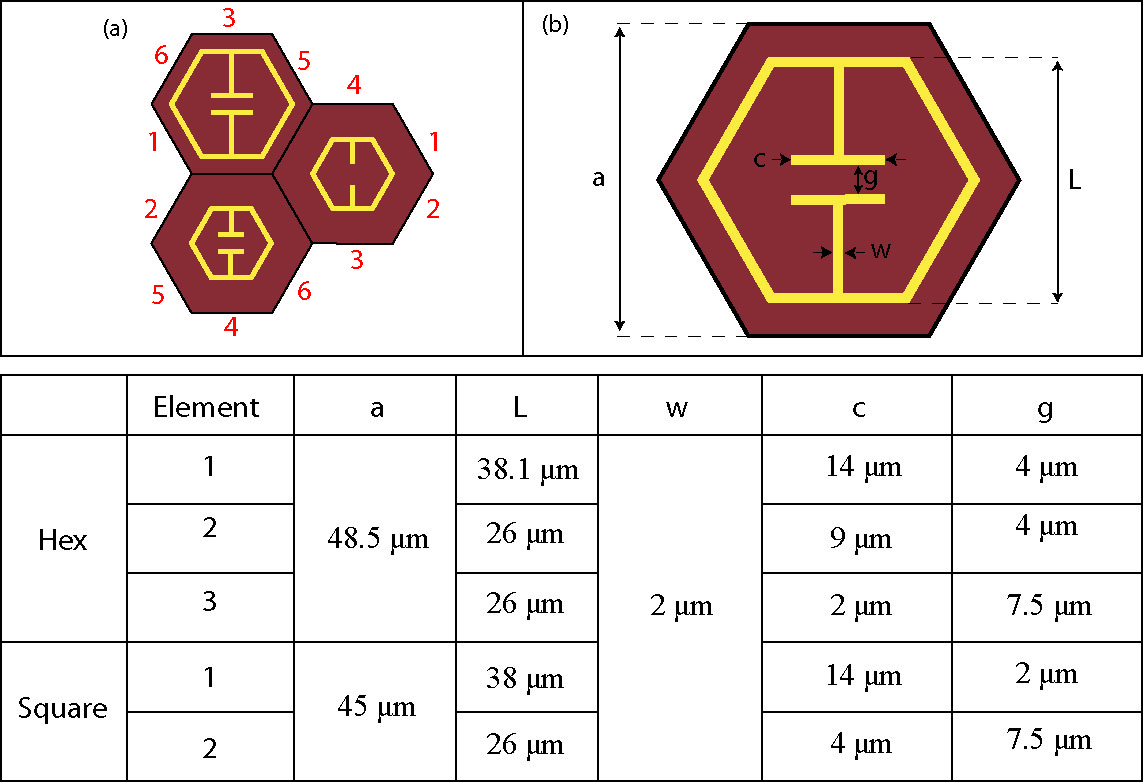}%
\caption{Simulated unit cell and a single primitive cell. (a) Diagram of simulated n=3 hexagonal MM. Also shown
are the assigned periodic boundary pairs used in simulation. (b) Single primitive cell with dimensions labeled.
The lower portion of the figure gives the values of the labeled parameters used in our designs.}
\label{fig2}%
\end{center}
\end{figure}

\section{Simulation}

In order to determine the correct dimensions and geometry for
resonant THz metamaterials consisting of multi-unit cells, we
simulated various structures of similar size to previous
work \cite{padillaPRB}. Metamaterials were computationally designed using a
commercial simulation package HFSS from Ansoft \cite{HFSS}. As a
general starting point for our structures an electric metamaterial was chosen due to its strong interaction with a
normally incident electromagnetic wave \cite{schurigAPL06}. For
metamaterials consisting of n=2 square unit cells (e.g. Fig.
\ref{fig4} (e)) we use simple
Perfect Electric Conductor (PEC) and Perfect Magnetic Conductor (PMC)
boundary conditions on the lateral faces of the unit cell. Waveguide ports on the other boundaries then
approximate propagation of a plane wave incident on the metamaterial
structure.

For structures consisting of n=3 or more different
primitive cells, we utilize a more amenable geometry ideal for
maximization of the electromagnetic response, as determined by Eq.
\ref{fill}. As such we use an underlying hexagonal unit cell, where
the electric metamaterials are then modified to maximize the filling
fraction, as shown in Fig. \ref{fig4}(a)-(d). Obviously for this
type of unit cell, the conventional boundary conditions described
above are unsuitable, and thus we use Periodic Boundary Conditions
(PBCs). A local coordinate system is created on each surface and the
field of one surface is mapped onto another. Assignment of these
boundary conditions is done by grouping the outer surfaces of the
model into pairs. For example, Fig. \ref{fig2}(a) shows the unit
cell for a n=3 metamaterial. Numbers around the perimeter denote a
particular surface and each is mapped to its pair. In this way we
are able to simulate a metamaterial infinite in extent in the
lateral directions. Plane wave ports on the front and back faces
then permit frequency dependent computational studies of the complex
S-parameters, i.e. transmission (T=$|S_{21}|^2$) and reflection
(R=$|S_{11}|^2$). Similar boundary mappings are used for the n=1
metamaterial structures shown in Fig. \ref{fig4} (a)-(c).

 \begin{figure}[ptb]
\begin{center}
\includegraphics[ width=5in,keepaspectratio=true]{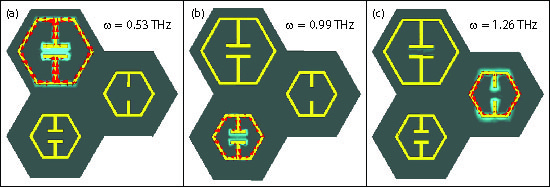}%
\caption{ Simulated electric field and surface current for n=3 hexagonal metamaterial(a) current density and
electric field plots at $\omega_1$ (b) current density and
electric field plots at $\omega_2$ (c) current density and
electric field plots at $\omega_3$ }
\label{fig3}%
\end{center}
\end{figure}

Simulated electric fields and current densities for the n=3
hexagonal metamaterials are shown in Fig. \ref{fig3}(a)-(c). We have
designed a metamaterial with three separate distinct structures, and
as such, we expect each sub-lattice of the metamaterial to be
resonant at different frequencies. Computational results shown in
Fig. \ref{fig3} (a)-(c) verify that this is the case. For example,
at $\omega_1\equiv$0.53 THz the metamaterial element in the top left
of the unit cell is resonant, as verified by the enhanced electric
field located within the gap, as well as the significant current
density - denoted by the red cones. The other two elements resonate at $\omega_2\equiv$0.99 THz and $\omega_3\equiv$1.26 THz
respectively. The electric field is plotted over the
entire unit cell, while current density only on the surface of each
metamaterial. These two parameters are normally out of phase by 90 degrees,
the current densities shown in Fig. \ref{fig3} have been shifted to
highlight the resonant response. This verifies that at each of the
frequencies $\omega_1$, $\omega_2$, and $\omega_3$ only a single
element is active, meaning that the combined metamaterial will
behave as if it was a single element structure over a frequency
range.

\begin{figure}[ptb]
\begin{center}
\includegraphics[width=5in,keepaspectratio=true]{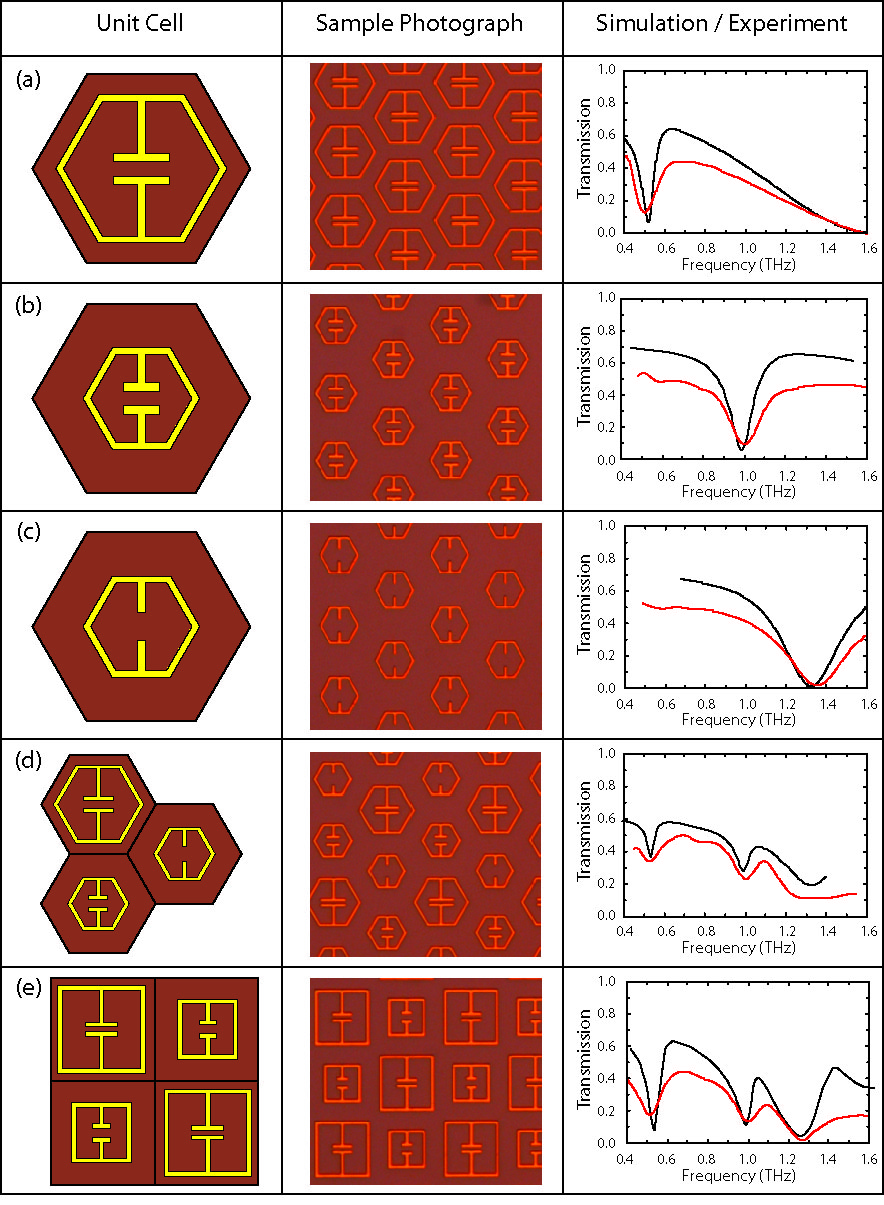}%
\caption{Simulation and experimental results for our different n=1,2, and 3 metamaterial designs.  The left column shows the unit cell models, images taken of the fabricated samples are in the middle column,
and plots of simulated (black curve) and experimental data (red curve) are shown in the right column.}
\label{fig4}%
\end{center}
\end{figure}

We now turn to the simulated transmission obtained for the various
metamaterial structures shown as the black curves in Fig.
\ref{fig4}. Each of the n=1 hexagonal metamaterials, shown in Fig.
\ref{fig4} (a)-(c) yield transmission minima at $\omega$=0.53 THz,
$\omega$=0.99 THz, and $\omega$=1.26 THz, respectively. Away from
resonance values of T = 60-70 percent are achieved, and minima at
resonance are below 10\%. Each of the unit cells shown in Fig.
\ref{fig4} (a)-(c) are used as primitive cells and combined into a
n=3 structure, as depicted in Fig. \ref{fig4}(d). Each of the
independent resonances are preserved in this combined metamaterial
structure, and we observe resonance at $\omega$=0.53 THz,
$\omega$=0.99 THz, and $\omega$=1.32 THz. As a verification of the
versatility and engineerability of metamaterials, we also design a
n=2 square checkerboard structure which also yields the same three
resonant frequencies, i.e. $\omega_1$, $\omega_2$, and $\omega_3$,
shown in Fig. \ref{fig4} (e). In this case for the highest frequency
resonance, we utilize the cut-wire response of the metamaterial,
rather than a distinct element\cite{padillaPRL}.

\section{Fabrication}

The metamaterials were fabricated at Boston University utilizing a standard lift off process. AZ5214e image reversal photoresist was spin-coated on HMDS-coated SI-GaAs wafer at 4,000rpm for 40s yielding a thickness of ~ 1.5 $\mu$m and baked on the hot plate at the temperature of 110$^{\circ}$C. The photoresist was exposed to UV light to pattern the metamaterial structures. A 150s hard bake was performed on a 120$^{\circ}$C hot plate, followed by flood-exposure for 60s. The wafer was then developed in the AZ 400K (1:4 diluted) for 15s to dissolve the resist where the metal should be deposited and rinsed in the DI water for 60s. A 10nm Ti layer was first deposited to improve the Au layer adhesion, and then a 200 nm-thick Au layer was evaporated using Sharon E-beam evaporator and then lifted off by rinsing in acetone for several minutes. Various images of the fabricated metamaterials
are shown in Fig. \ref{fig4}.

\section{Experiment}

All metamaterial samples were characterized at Boston College using
a Fourier Transform Infrared (FTIR) spectrometer. THz radiation from
a mercury arc-lamp, was polarized before impinging on MM samples at
normal incidence. The transmitted THz was then refocused on a liquid
helium silicon bolometer and recorded as a function of mirror step
position of the FTIR. Etalons due to multiple reflections within the
GaAs substrate were removed and the modified interferrogram was
Fourier transformed to obtain the sample spectrum. A similar
procedure was performed for a reference with an open channel and
division of the sample and reference spectra resulted in the
frequency dependent absolute value transmission T($\omega$).

In the right column of Fig. \ref{fig4} we show experimental
transmission (red curves) of the metamaterials. Each metamaterial
sample yields a resonant response in the THz frequency range. For
example, the sample shown in Fig. \ref{fig4}(a) yields low frequency
transmission values near 50 percent before a minimum occurs,
yielding a value of 12 percent at 0.5 THz, before recovering back to
relatively high values. At higher frequencies, the curve tends toward T=0, as higher order resonances occur.
Similar behavior is observed for the other n=1 hexagonal
metamaterial elements, with resonances at $\omega$=1.0 THz (Fig.
\ref{fig4}(b)), and $\omega$=1.35 THz (Fig. \ref{fig4}(c)). When all
of the three individual metamaterial elements are combined into a
single structure (Fig. \ref{fig4}(d)), resonances observed in the
individual hexagonal metamaterials are preserved with very little
change in frequency location.

Electromagnetic responses to that of the n=3 hexagonal
structure, can be designed and obtained for a n=2 square
metamaterial. In Fig. \ref{fig4}(e) the red curves show T($\omega$)
for the square n=2 checkerboard metamaterial. Transmission minima
occur at $\omega$=0.51 THz, $\omega$=1.0 THz, and $\omega$=1.26
THz, in great agreement with that obtained for the n=3 hexagonal MM.
Notice in this case, that although transmission maxima are similar,
i.e. T$\sim$40-50\%, that oscillator strengths are greater for the
square n=2 lattice compared to that of the n=3 hexagonal
metamaterial. In the former, minima in T($\omega$) are all 20\% or
lower, whereas in the latter they range between 20 and 40\%. This
can be understood directly from Eq. 2, since the oscillator strength
$F_m$ is inversely proportional to the number of elements (n).

\section{Discussion}
\begin{figure}[ptb]
\begin{center}
\includegraphics[width=5in,keepaspectratio=true]{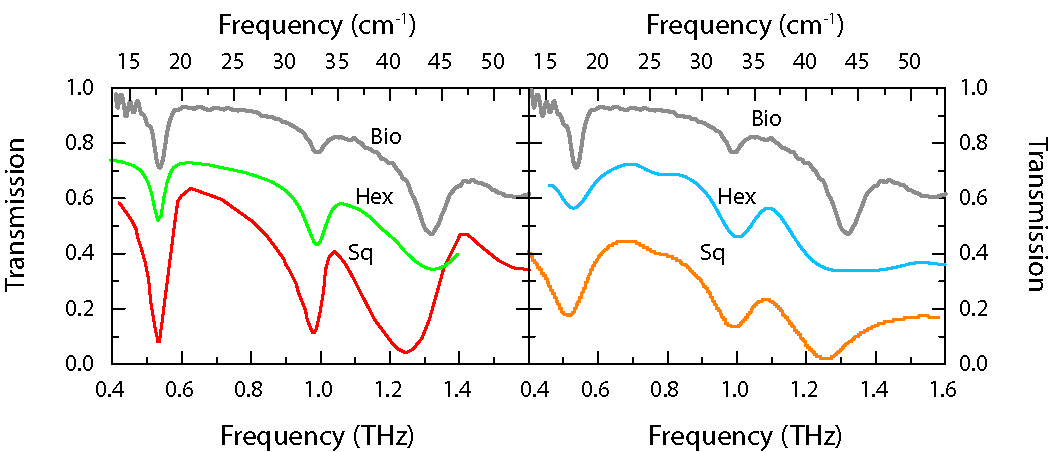}%
\caption{Computational (left) and experimental (right) measurements
of the n=3 hexagonal metamaterial, and the n=2 square checkerboard
metamaterial, compared to experimental measurements of the molecule
biotin. Simulated and experimental transmission spectra of the
hexagonal metamaterial has been shifted up by 20\% for clarity and
T($\omega$) of biotin is in arbitrary units.}
\label{fig5}%
\end{center}
\end{figure}

We now turn toward highlighting some potential applications of these
planar wallpaper group metamaterials. The frequencies of roughly
$\omega$=0.5, 1, 1.25 THz have been specifically chosen in order to
investigate the possibility of metamaterials to act as
bio-detectors, or as a means of electromagnetic mimicry. For
example, the resonant frequencies of our metamaterials were designed
to coincide with the electromagnetic resonances of the molecule
biotin, (C$_{10}$H$_{16}$N$_2$O$_3$S) also known as vitamin H or
B$_7$. These natural resonances result from rotational, twisting, or
vibrational modes and typically occur in the microwave through THz
and far infrared ranges of the electromagnetic
spectrum \cite{biotin,chem}. As such, biotin is
an excellent molecule to benchmark possible bio-detection
methods at THz frequencies. The grey curves of Fig. \ref{fig5} show the measured THz
transmission spectra of biotin, (powdered form), which reveal three
resonances at $\omega$=0.5 THz, $\omega$=1.0 THz, and $\omega$=1.26
THz, consistent with previous data ~\cite{biotin}.

Most proposed ideas for bio-detection utilizing metamaterials
involve using the enhanced electric fields near or within the gap of
split-rings~\cite{detector1, detector2}. In this scheme, when the
desired particle comes into the vicinity of the gap there is a
change in the resonant frequency ($\omega_0$) of the metamaterial,
due to the sensitivity on the capacitance ($C$), i.e.
$\omega_0\sim1/\sqrt{C}$. In-fact the discovery publication
~\cite{pendry99} on split-ring resonators highlighted the enhanced
field within the gap and suggested its use as a means to enhance
non-linear phenomena. Indeed experimental results at THz frequencies
have verified the extent to which features of the metamaterial
resonance can be controlled by varying gap properties ~\cite{Padilla2006,Chen2006}. However it should be mentioned that the
above methods ~\cite{detector1,detector2} are prone to contaminants
which may cause false positives in identification. For example, \textit{any
dielectric which lies in vicinity of the gap} has the ability to
modify the resonant frequency. Thus particles as common as - say -
dust and pollen will be severely problematic.

However, it may be possible to perform bio / chemical detection with metamaterials by
non-linear processes in analogy with surface enhanced Raman scattering (SERS), or more closely, surface enhanced infrared absorption spectroscopy (SEIRA) \cite{siera}. This idea is based on the significant field enhancement which occurs within the gaps of SRRs. Simulations indicate that within the gaps of the metamaterial the on-resonance electric field for the structures shown in Fig. \ref{fig3} is of
order 10$^6$ - 10$^7$ V$\cdot$m$^{-1}$ when illuminated with a power
of 10$^4$ W$\cdot$cm$^{-2}$. Knowing the specific resonance frequencies of a particular substance is the first step for
building such a device. Identification of a molecular hazard could
then be achieved by examining how the metamaterial and molecule,
which are both resonant at the same frequency, behave upon binding of the molecule within the SRR gap. The response of the SRR array (with biotin occupying the gap) will yield an enormous, perhaps
non-linear response, thus increasing the sensitivity by as much as
10$^6$ - typical with driven resonances or overlapping resonant
phenomena  ~\cite{overlap1,overlap2}. In addition, previous work has shown that the THz
resonances of biotin have anharmonicity constants ranging from
10$^{-4}$ to 10$^{-3}$ ~\cite{biotin}. Thus another possible identification method
involves detection of the sub or super-harmonic modes
created upon excitation. In short, nonlinearities associated with overlapping
resonant metamaterials and molecular responses, through engineering metamaterials to a bio or
chemical hazard of interest, will provide an interesting approach beyond simple dielectric induced resonance shifts.

In order to negate false positives, proper identification of a
material may require more than a single frequency match. As demonstrated throughout this paper a salient
feature of metamaterials is the ability to create a structure that
yields resonances at several frequencies. This
scheme, often called ``fingerprinting" would then utilize several points of
comparison to ensure accuracy. An initial study has been carried out
and in Fig. \ref{fig5} we show a comparison of T($\omega$) of biotin
to various metamaterial structures. Both the n=2 square MM and the
n=3 hexagonal MM, achieve a good match to biotin, right panels. Thus
when biotin lies in proximity to various gaps of the structure, the
MM resonances (when excited by THz radiation) will overlap with the
THz resonances of biotin. Future detection could be enhanced by
patterning specific binding agents to metamaterial gaps. In the case
of biotin the protein avidin could be used, which has an affinity of
order 10$^{15}$ M$^{-1}$. ~\cite{zhang} Whatever the desired molecule
or agent of interest, site specific binding can yield reductions in
false positives, and may help to increase the sensitivity of
detection.

Another potential application of metamaterials demonstrated here is
that of electromagnetic signature control. For example - from the
viewpoint of interrogating THz radiation - both our n=2 square and
n=3 metamaterials look nearly identical to biotin. Thus utilizing the
methods shown here for the design and fabrication of metamaterials,
one could construct materials, or the surfaces of structures, to
mimic another substance. Further, recently it has been shown
possible to fashion metamaterials as perfect absorbers, both at
microwave ~\cite{landy} and THz ~\cite{taoOpEx08}. Kirchhoff's law of
thermal radiation dictates that the emission spectra of such a
metamaterial should then be, at thermal equilibrium, equal to that
of the absorption spectrum. Thus, with multi-unit cell ideas here,
combined with perfect absorber techniques, one can design a specific
emission spectra for an object at a given temperature, i.e. engineer its blackbody spectra.

\section{Conclusion}

Our results verify that multiple element metamaterials can be
successfully designed, fabricated, and measured at THz frequencies.
Two different structures were used to obtain resonances at roughly
$\omega$=0.5 THz, $\omega$=1 THz, and $\omega$=1.25 THz. We have
constructed a metamaterial structure composed of three hexagonal
sub-elements that mimic the resonant behavior seen in the molecule
biotin. The simulated and the experimentally measured transmission
were also in good agreement. Hexagonal structures opens up potential
new methods for creating multiple resonator metamaterials. Such multi-resonator metamaterials may provide
a quick method for fingerprinting and detection of chemical and
biological agents based on nonlinearities associated with the electric
field enhancement within the capacitive gaps, or as materials to facilitate electromagnetic signature control.\\

\noindent \textbf{Acknowledgement}\\

\noindent WJP and CMB acknowledge support from the Office of Naval Research (ONR), grant N000140710819. RDA and XZ acknowledge support from NSF EECS 0802036. The authors would also like to thank the Photonics Center at Boston University for technical support.

\end{document}